# Relaxation and Creep in Twist and Flexure

*V. Kobelev[1]*

**Abstract**

The aim of the paper is to derive the exact analytical expressions for torsion and bending creep of rods with the Norton-Bailey, Garofalo and Naumenko-Altenbach-Gorash constitutive models. These simple constitutive models, for example, the time- and strain-hardening constitutive equations, were based on adaptations for time-varying stress of equally simple models for the secondary creep stage from constant load/stress uniaxial tests where minimum creep rate is constant. The analytical solution is studied for Norton-Bailey and Garofalo laws in uniaxial states of stress. The most common secondary creep constitutive model has been the Norton-Bailey Law which gives a power law relationship between minimum creep rate and (constant) stress. The distinctive mathematical properties of the power law allowed the development of analytical methods, many of which can be found in high temperature design codes. The results of creep simulation are applied to practically important problem of engineering, namely for simulation of creep and relaxation of helical and disk springs. The exact analytical expressions giving the torque and bending moment as a function of the time were derived.

[1] kobelev@imr.mb.uni-siegen.de

# 1. Introduction

Stress analysis for creep has a long history in engineering mechanics driven by the requests of design for elevated temperature. The examples of high loaded elements of machinery deliver the springs made of steel. Steel springs are the typical energy storing elements of valve train in engines, clutches and automatic transmissions of cars. The coned-disk spring, Belleville spring or cupped spring washer, or Belleville washers are typically used as springs, or to apply a pre-load or flexible quality to a bolted joint or bearing. As the basic properties of Belleville washers include high fatigue life, better space utilization, low creep tendency, high load capacity with a small spring deflection. The physical phenomenon with metal springs is that at stress below the yield strength of the material a slow inelastic deformation take place. In the spring branch this is called creep when a spring under constant load loose length and it is called relaxation when a spring under constant compression lose load. The creep and relaxation rates depend on the temperature, the stress in the metal, the yield strength and the time. Increased temperature, stress and time also increase the creep and relaxation rates. Especially the temperature and stress have a major influence. The precise creep description is essentially important for correct dimensioning of springs.

The springs store the elastic energy either by means of bending or torsion. Respectively, in material dominates either uniaxial or pure shear stress state. Thus, the essential task is the derivation of the exact closed form expressions for torsion and bending creep for isotropic materials, which obey the commonly accepted Norton-Bailey, Garofalo and Naumenko-Altenbach-Gorash constitutive laws. These basic constitutive models were based on the time- and strain-hardening constitutive equations for time-varying stress. The models adequately describe the secondary creep stage from constant load/stress uniaxial tests where creep rate is nearly constant. Among others, the most widespread secondary creep constitutive model has been the Norton-Bailey law which provides a power law relationship between creep rate and stress. The distinctive mathematical properties of the power law allowed the development of analytical methods, many of which can be found in high temperature design codes. The constitutive models and the solution methods for creep problems are discussed in (Kassner [1]). A summary of creep laws for common engineering materials is provided in [2,3]. The results of creep simulation are applied to practically important problem of engineering, namely for simulation of creep and relaxation of helical and disk springs.

The helical springs are made up of a wire coiled in the form of a helix and are primarily intended for compressive or tensile loads. The cross-section of the wire from which the spring is made may be circular, square or rectangular. Helical compression springs have applications to resist applied compression forces. A formal technique was developed in [4] to predict the stress relaxation for compression and torsion springs. The technique uses uniaxial tensile-generated stress-relaxation data for spring wires. Based on the tension-induced stress-relaxation data, the technique was applied to compression springs, where shear stress dominates in predicting the stress relaxation. The shear stress-strain curve is first constructed based on the uniaxial tension stress-strain curve. Next, an understanding is established that stress relaxation is a phenomenon in which part of the elastic strain responsible for the initial stress is replaced by creep strain.

# 2. Constitutive equations for creep

### 2.1. Tensorial generalization of creep laws

The creep component of strain rate is defined by material specific creep law. In this article we adopt, following the common procedure (Betten [5]), an isotropic stress function

(1) $\dot{\varepsilon}'_{ij} = \frac{3\sigma'_{ij}}{2\sigma_{eff}} F(\sigma_{eff}, t)$.

The special case of incompressible behavior of material ($\dot{\varepsilon}_{kk} = 0$) is assumed hereafter.
In the Equation (1) the following notations are used:

$\dot{\varepsilon}'_{ij} = \dot{\varepsilon}_{ij} - \frac{1}{3}\dot{\varepsilon}_{kk}\delta_{ij}$  is the deviatoric component of creep strain rate,

$\sigma'_{ij} = \sigma_{ij} - \frac{1}{3}\sigma_{kk}\delta_{ij}$  is the deviatoric component of stress,

$\sigma_{eff} = \sqrt{\frac{3}{2}\sigma'_{ij}\sigma'_{ij}} = \sqrt{3J'_2}$  is the Mises equivalent stress,

$J'_2 = \frac{1}{2}\sigma'_{ij}\sigma'_{ij}$  is the second invariant of the stress tensor.

We derive the expressions for strain rate for uniaxial and shear stress states for the definite representations of stress function.

### 2.2. Norton-Bailey Law

Firstly, consider Norton-Bailey law [6]. The isotropic stress function reads in this case

(2) $F(\sigma_{eff}, t) = \bar{\varepsilon}\left(\frac{t}{\bar{t}}\right)^{k-1}\left(\frac{\sigma_{eff}}{\bar{\sigma}}\right)^{m+1}$,

where $\bar{\varepsilon}$, $\bar{\sigma}$, $\bar{t}$, $m$ and $k$ are the experimental constants.
There is only non-vanishing component of stress tensor ($\sigma_{11} = \sigma$) for the uniaxial stress state. Correspondingly, the non-vanishing components of strain rate are

$\dot{\varepsilon} \equiv \dot{\varepsilon}_{11} = -2\dot{\varepsilon}_{22} = -2\dot{\varepsilon}_{33}$,

where

$\dot{\varepsilon} = \bar{\varepsilon}\left(\frac{t}{\bar{t}}\right)^{k-1}\left(\frac{\sigma}{\bar{\sigma}}\right)^{m+1}$.

For brevity of equations, we introduce the material constant

$c_\sigma = \frac{\bar{\varepsilon}}{\bar{t}^{k-1}\bar{\sigma}^{m+1}}$.

With this constant the dependence of uni-axial strain rate upon stress reads

(3) $\dot{\varepsilon} = c_\sigma t^{k+1}\sigma^{m+1}$.

For pure shear stress state ($\sigma_{12} = \sigma_{21} = \tau$) the non-vanishing components of deformation rate are

$\dot{\varepsilon}_{12} = \dot{\varepsilon}_{21} \equiv \frac{\dot{\gamma}}{2} = \frac{1}{2}\bar{\gamma}\left(\frac{t}{\bar{t}}\right)^{k-1}\left(\frac{\tau}{\bar{\tau}}\right)^{m+1}$,

where

$\bar{\gamma} = \sqrt{3}\bar{\varepsilon}$.

With the creep constant for shear strain

$$c_\tau = \frac{\bar{\gamma}}{\bar{t}^{k-1}\bar{\tau}^{m+1}}$$

and equivalent shear stress

$$\bar{\tau} = \frac{\bar{\sigma}}{\sqrt{3}}$$

the Norton-Bailey creep law for pure shear deformation reduces to

(4) $\quad \dot{\gamma} = c_\tau t^{k-1} \tau^{m+1}$ .

There is a simple relation between the constants in Equations (3) and (4):

$$c_\tau = 3^{m/2+1} c_\sigma .$$

### 2.3. Garofalo creep Law

Secondly, the isotropic stress function for Garofalo creep law [7,8] could be represented as

(5) $\quad F(\sigma_{eff}, t) = \bar{\varepsilon}\left(\frac{t}{\bar{t}}\right)^{k-1} \sinh\left(\frac{\sigma_{eff}}{\bar{\sigma}}\right)$ .

For the uniaxial stress state the deformation rate reads

$$\dot{\varepsilon} = \bar{\varepsilon}\left(\frac{t}{\bar{t}}\right)^{k-1} \sinh\left(\frac{\sigma}{\bar{\sigma}}\right)$$

or

(6) $\quad \dot{\varepsilon} \equiv \dot{\varepsilon}_{11} = c_\sigma t^{k-1} \sinh\left(\frac{\sigma}{\bar{\sigma}}\right)$

Here

$$c_\sigma = \frac{\bar{\varepsilon}}{\bar{t}^{k-1}}$$

is the creep constant for uni-axial strain.
For pure shear stress state the deformation rate reads

$$\dot{\varepsilon}_{12} = \dot{\varepsilon}_{21} = \frac{1}{2}\bar{\gamma}\left(\frac{t}{\bar{t}}\right)^{k-1} \sinh\left(\frac{\sqrt{3}\tau}{\bar{\sigma}}\right) .$$

Finally, the shear strain rate according to Garofalo creep law is

(7) $\quad \dot{\gamma} \equiv 2\dot{\varepsilon}_{12} = c_\tau t^{k-1} \sinh\left(\frac{\tau}{\bar{\tau}}\right)$

with the corresponding constant

$$c_\tau = \frac{\sqrt{3}\bar{\varepsilon}}{\bar{t}^{k-1}} \equiv \frac{\bar{\gamma}}{\bar{t}^{k-1}} \equiv \sqrt{3} c_\sigma .$$

### 2.4. Naumenko-Altenbach-Gorash Law

Thirdly, the isotropic stress function for Naumenko-Altenbach-Gorash creep law [9] is

$$(8) \quad F(\sigma_{eff}, t) = \bar{\varepsilon} \cdot \left[ \frac{\sigma_{eff}}{\bar{\sigma}} + \left(\frac{\sigma_{eff}}{\bar{\sigma}}\right)^{m+1} \right].$$

For the uniaxial stress state the strain rate reads

$$(9) \quad \dot{\varepsilon} \equiv \dot{\varepsilon}_{11} = \bar{\varepsilon} \cdot \left[ \frac{\sigma}{\bar{\sigma}} + \left(\frac{\sigma}{\bar{\sigma}}\right)^{m+1} \right] = -2\dot{\varepsilon}_{22} = -2\dot{\varepsilon}_{33}.$$

For pure shear stress state the shear deformation rate reduces to

$$(10) \quad \dot{\gamma} \equiv 2\dot{\varepsilon}_{12} = \bar{\gamma} \cdot \left[ \frac{\tau}{\bar{\tau}} + \left(\frac{\tau}{\bar{\tau}}\right)^{m+1} \right].$$

For the creep laws (1)-(10) the closed form solutions of basic creep problems are derived., Numerical values for the creep constants $\bar{\varepsilon}$, $\bar{\gamma}$, $\bar{\sigma}$, $\bar{\tau}$ are apparently different for diverse creep laws.

## 3. Creep and Relaxation of helical compression springs

### 3.1. Basic constitutive equations for relaxation in torsion

The deformation of body during relaxation does not alter, but the stress gradually reduces. Consider the relaxation problem for a rod with circular cross-section under the constant twist. Let $\tau(r,t)$ is shear stress in the cross-section of rod. The total shear strain in any instant of the time is $\gamma(r,t)$, is the sum of the elastic and the creep components of shear strain:

$$(11) \quad \gamma = \gamma_e + \gamma_c.$$

The creep component of shear strain is $\gamma_c(r,t)$. The elastic component of shear strain is

$$(12) \quad \gamma_e = \tau/G,$$

where $G$ is the shear modulus.

Firstly, in this Article we investigate the creep for the total deformation that remains constant in time. Thus, we consider the total strain $\gamma_0(r)$ as function of radius only. However, the elastic and the creep components of strain are the functions as well of radius and of time, such that:

$$(13) \quad \gamma(r,t) = \gamma_e(r,t) + \gamma_c(r,t) \equiv \gamma_0(r).$$

The time differentiation of (3) leads to the differential equation for elastic and creep strain rates:

$$(14) \quad \dot{\gamma}(r,t) = \dot{\gamma}_e(r,t) + \dot{\gamma}_c(r,t) \equiv 0,$$

where dot denotes the time derivative.
The differentiation of the equation (12) over time delivers the elastic component of strain rate

$$(15) \quad \dot{\gamma}_e = \dot{\tau}/G.$$

### 3.2. Torque relaxation for the materials, that obeys Norton-Bailey Law

At first, we assume the Norton-Bailey law for the state of shear stress [10]

$$(16) \quad \dot{\gamma}_c(r,t) = c_\tau t^{k-1} \tau^{m+1},$$

The substitution of material law (5) in Equation (14) results in the ordinary nonlinear differential equation of the first order for total shear stress $\tau(r,t)$:

(17) $\quad \dfrac{\dot{\tau}}{G} + c_\tau t^{k-1} \tau^{m+1} = 0$.

The initial condition for the equation (17) presumes the pure elastic shear stress in the initial moment $t = 0$:

(18) $\quad \tau(r, t = 0) = \tau_0(r)$.

The shear stresses in the moment $t = 0$ for the rod with circular cross-section are

$$\tau_0(r) = G\theta r,$$

where $\theta$ is the twist angle per unit length. The torque at the moment $t = 0$ is

$$M_T^0 = \frac{1}{2} G \pi \theta R^4.$$

The solution of the ordinary differential equation (17) with initial condition (18) delivers the shear stress over the cross-section of the twisted rod as the function of time and radius:

(19) $\quad \tau(r,t) = \left[ \tau_0^{-m}(r) + \dfrac{c_\tau G m t^k}{k} \right]^{-1/m} \equiv G\theta \left[ \dfrac{1}{r^m} + \dfrac{c_\tau G^{m+1} \theta^m m}{k} t^k \right]^{-1/m}$.

The couple as the function of time is

$$M_T(t) = 2\pi \int_0^R r^2 \tau(r,t) dr.$$

With the expression for total shear stress (19) we can calculate the couple

(20) $\quad M_T(t) = 2\pi G \theta \int_0^R r^2 \left[ \dfrac{1}{r^m} + \dfrac{c_\tau G^{m+1} \theta^m m}{k} t^k \right]^{-1/m} dr$

For evaluation of the integral (20) the formula for $J_p(a,m;X)$ from Appendix B is applied for the case $p = 2$. The integral could be expressed in terms of hypergeometric function [11]

(21) $\quad M_T(t) = 2\pi G \theta J_p\left( \dfrac{c_\tau \theta^m G^{m+1} m t^k}{k}, m; R \right) =$
$\quad = {}_2F_1\left( \dfrac{4}{m}, \dfrac{1}{m}; \dfrac{4+m}{m}; -\dfrac{c_\tau \theta^m G^{m+1} m t^k}{k} R^m \right) M_T^0.$

### 3.3. Torque relaxation for the material, that obeys Garofalo Law

At second, we presume the Garofalo law for uniaxial state of stress

(22) $\quad \dot{\gamma}_r(r,t) = c_\tau t^{k-1} \sinh\left( \dfrac{\tau}{\bar{\tau}} \right)$.

The solution of the differential equation (14) with initial condition (18) for the Garofalo creep law (22) reads

   

$$(23) \quad \tau(r,t) = \bar{\tau} \ln \left\{ \tanh \left[ \frac{Gc_\tau}{2k\bar{\tau}} t^k + \text{arctanh}\left( \exp\left( \frac{Gr\theta}{\bar{\tau}} \right) \right) \right] \right\}.$$

Using formula for $I_2(a,b;X)$ from the Appendix A, the time dependent torque could be expressed in terms of polylogarithms [12]

$$(24) \quad M_T(t) = 2\pi \int_0^R r^2 \tau(r,t) dr = 2\pi \bar{\tau} I_2\left( \frac{Gc_\tau}{2k\bar{\tau}} t^k, \frac{G\theta}{\bar{\tau}}; R \right).$$

### 3.4. Torque relaxation for the material, that obeys Naumenko-Altenbach-Gorash

At third, we apply the Naumenko-Altenbach-Gorash law for the state of shear stress

$$(25) \quad \frac{\dot{\gamma}_r(r,t)}{\bar{\gamma}} = \frac{\tau}{\bar{\tau}} + \left( \frac{\tau}{\bar{\tau}} \right)^{m+1},$$

The substitution of material law (25) in Eq. (14) leads to the ordinary differential equation for total shear stress

$$(26) \quad \frac{\dot{\tau}}{G} + \bar{\gamma}\left[ \frac{\tau}{\bar{\tau}} + \left( \frac{\tau}{\bar{\tau}} \right)^{m+1} \right] = 0.$$

The solution of the differential equation (26) with initial condition (18) delivers the shear stress over the cross-section of the twisted rod

$$(27) \quad \tau(r,t) = \frac{r}{\left( r^m \frac{m\bar{\gamma}G\xi - \tau_c}{\bar{\tau}^{m+1}} + \frac{\xi}{G^m \theta^m} \right)^{1/m}}$$

with

$$\xi = \exp\left( \frac{m\bar{\gamma}G}{\bar{\tau}} t \right).$$

For evaluation the formula for $J_p(a,m;X)$ from Appendix B is applied for the case $p=2$. Relaxation of torque could be expressed again in terms of hypergeometric function:

$$(28) \quad M_T(t) = {}_2F_1\left( \frac{4}{m}, \frac{1}{m}; \frac{4+m}{m}; \frac{(\theta GR)^m}{\xi \bar{\tau}^{m+1}}(\bar{\tau} - \bar{\gamma}\xi Gm) \right) \xi^{-1/m} M_T^0.$$

### 3.5. Relaxation and creep of helical compression springs

Compression springs are made of an elastic wire material formed into the shape of a helix, which returns to its natural length when unloaded. Compression springs can be commonly referred to as a coil spring or a helical spring. Coil springs are a mechanical device which is typically used to store energy and subsequently release it to absorb shock, or to maintain a force between contacting surfaces. Compression springs are made of an elastic wire material formed into the shape of a helix (which is why they are also called helical springs), which returns to its natural length when unloaded. Compression springs can be commonly referred to as a coil spring or a helical spring. Coil springs are a mechanical device which is typically used to store energy and subsequently release it to absorb shock, or to maintain a force be-tween contacting surfaces. The major stresses produced in conical and volute springs are also shear stresses due to twisting.

Consider the helical spring with mean helix diameter $D$, wire diameter $d$ and active coils number $n_a$. The force $P_z(t)$ applied to a spring that causes a deflection $s$ is

$$P_z(t) = 2M_T(t)/D.$$

The relaxation of helical springs occurs, when the compressed length of the spring remains constant. In this circumstance the spring force $P_z(t)$ reduces over time.

On the contrary, the creep of helical springs occurs, when the spring force of the spring remains constant. In this situation the compressed length continuously reduces over time and the deflection increases [13].

### 3.6. Relaxation of helical compression springs

For the solution of relaxation problem we apply the results of relaxation problem for twisted rod with the circular cross-section. If the material of helical springs obeys Norton-Bailey law, the spring load as the function of time is

$$(29) \quad P_z(t) = {}_2F_1\left(\frac{4}{m}, \frac{1}{m}; \frac{4+m}{m}; -\frac{c_\tau d^m \theta^m G^{m+1} m t^k}{2^m k}\right) P_z^0,$$

where the twist angle per unit length is

$$\theta = \frac{2s}{\pi n_a D^2}.$$

The spring force at initial moment $t=0$ due to pure elastic deformation in (29) is

$$P_z^0 = \frac{2M_T^0}{D} = \frac{\pi G d^4}{16 D} \theta = \frac{G d^4}{8 n_a D^3} s.$$

The expression for shear stress on the outer surface of wire at $t=0$ is the function of axial deflection

$$\tau_0 = \frac{Gd}{\pi D^2 n_a} s.$$

With the value $\tau_0$ the expression (29) reduces to

$$(30) \quad P_z(t) = {}_2F_1\left(\frac{4}{m}, \frac{1}{m}; \frac{4+m}{m}; -\frac{c_\tau G m}{k} \tau_0^k t^k\right) P_z^0.$$

The further simplification expressions for spring load as the function of time is achieved for certain exponents in Norton-Bailey law:

$$P_z(t) = \frac{4k}{3} \frac{\left(1 + c_\tau G t^4 \tau_0^4 / k\right)^{3/4} - 1}{c_\tau G t^4 \tau_0^4} P_z^0 \qquad \text{for } m = 4$$

and

$$P_z(t) = \frac{2k}{3} \frac{k + \left(c_\tau G t^2 \tau_0^2 - k\right)\sqrt{1 + 2 c_\tau G t^2 \tau_0^2 / k}}{c_\tau^2 G^2 t^4 \tau_0^4} P_z^0 \qquad \text{for } m = 2.$$

The relaxation of spring load over time demonstrates the dimensionless relaxation function:

$$\Phi = \frac{P_z(t)}{P_z^0} = {}_2F_1\left(\frac{4}{m}, \frac{1}{m}; \frac{4+m}{m}; -\frac{c_\tau Gm}{k}\tau_0^k t^k\right).$$

On the Fig.1 the function $\Phi$ is plotted for different twist angles per unit length. For illustration of the procedure, we perform all calculations with Norton-Bailey law and the following set of material constants

$G = 79.74\,GPa,$   $E = 200\,GPa,$
$m = 4,$   $k = 1,$
$\bar{\varepsilon} = 10^{-24}\,\sec^{-1},$   $\bar{\gamma} = \sqrt{3}\bar{\varepsilon} = 1.73 \cdot 10^{-24}\,\sec^{-1},$
$\bar{\sigma} = 2000\,MPa,$   $\bar{\tau} = \bar{\sigma}/\sqrt{3} = 1154\,MPa.$

The three graphs of the function $\Phi$ are shown on the figure for twist angles per unit length ($\theta = 0.1; 0.2; 0.3$). It means that the higher twist angle per unit length, the quickly occurs the relaxation of torque. This apparently happens because the strain rate is higher for higher stresses.

On the Fig.2 the function $\Phi$ is plotted for the same twist angle per unit length ($\theta = 0.1$), but for the different exponents ($m = 3.0; m = 4.0; m = 5.0; m = 6.0$). The rest of material parameters remain the same as in the previous example. The Fig.2 obviously demonstrates, that the higher the exponent, the higher the relaxation rate.

Similar analytical procedure could be applied for Naumenko-Altenbach-Gorash creep law as well.

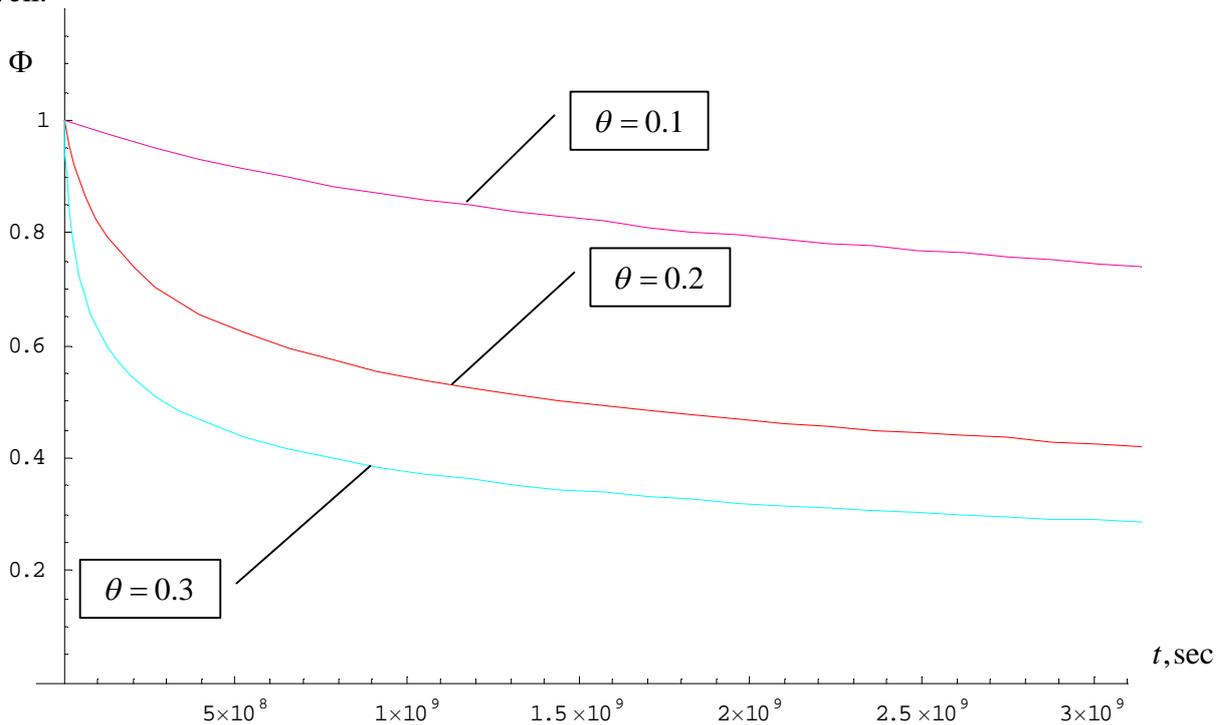

**Fig. 1 Relaxation function $\Phi$ for different twist rates**

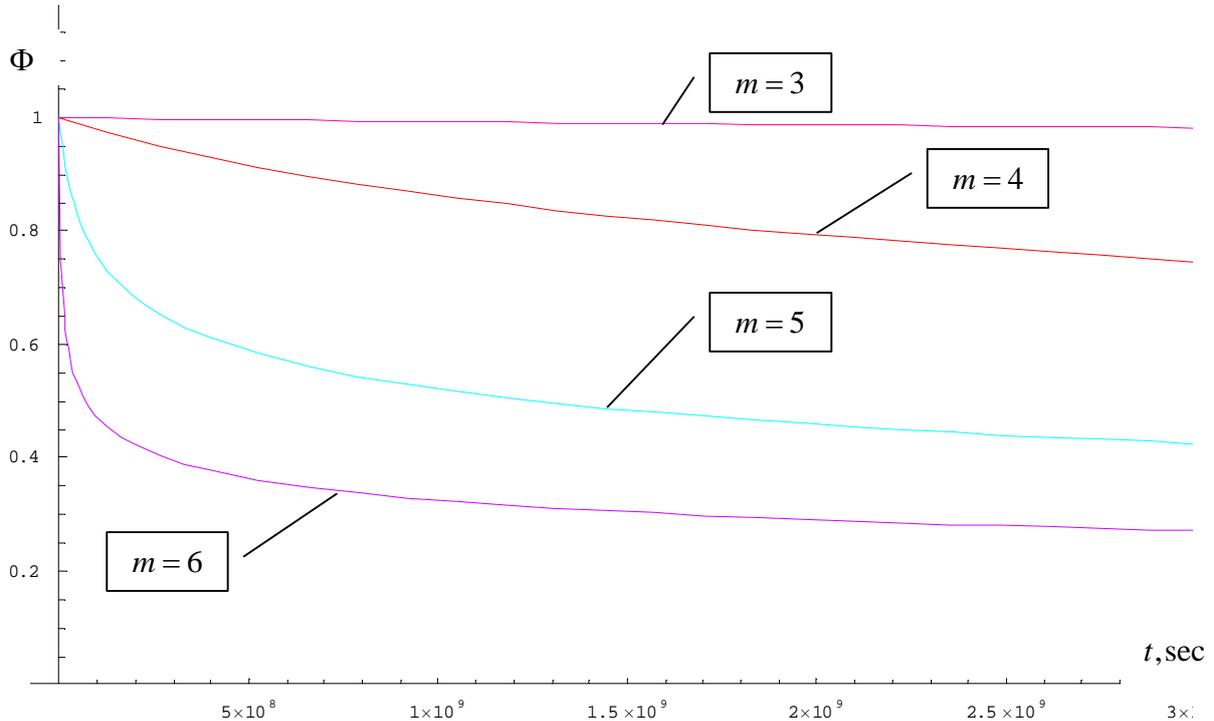

**Fig. 2** Relaxation function $\Phi$ for twist rate $\theta = 0.1$ and different creep exponents

### 3.7. Creep of helical compression springs

During the creep deformation of springs the spring force

$$P_z^0 = 2M_T^0 / D$$

remains constant over time. Correspondingly, the spring length reduces with time. The length decrease rate over time is $\dot{s}$. In the sector of circular wire with radius $r$ the shear strain rate is

$$\dot{\gamma}(t,r) = \dot{\theta} r = \frac{2\dot{s}r}{\pi n_a D^2}$$

for $0 < r < d/2$.

According to Norton-Bailey law (4) the shear stress due to creep is

$$\tau = \left(\frac{\dot{\gamma}}{c_\tau t^{k-1}}\right)^{\frac{1}{m+1}} = \left(\frac{2\dot{s}r}{c_\tau t^{k-1} \pi n_a D^2}\right)^{\frac{1}{m+1}}.$$

Performing the integration over the area of wire we get the moment due to creep

$$M_T^0 = 2\pi \int_0^R r^2 \tau(r,t)dr = 2\pi \frac{m+1}{4+3m} \left(\frac{2R^{4+3m}\dot{s}}{t^{k-1}\pi D^2 n_a c_\tau}\right)^{\frac{1}{1+m}} = \frac{\pi}{4} \frac{m+1}{4+3m} \left(\frac{2d^{4+3m}\dot{s}}{t^{k-1}\pi D^2 n_a c_\tau}\right)^{\frac{1}{1+m}}.$$

Solution of this equation delivers the spring length reduction rate for constant spring force $P_z^0$ as the function of time

(31) $$\dot{s} = \frac{t^{k-1}\pi D^2 n_a c_\tau}{2d^{4+3m}}\left(\frac{4M_T^0}{\pi}\frac{4+3m}{m+1}\right)^{m+1} = \frac{t^{k-1}\pi D^2 n_a c_\tau}{2d^{4+3m}}\left(\frac{4DP_z^0}{2\pi}\frac{4+3m}{m+1}\right)^{m+1}.$$

## 4.     Creep and Relaxation in the case of bending

### 4.1. Basic constitutive equations for relaxation in bending

Consider the problem of stress relaxation in the pure bending of a rectangular cross-section ($B \times H$) beam. In the applied Euler-Bernoulli theory of slender beams subjected to a bending moment $M_B$, a major assumption is that 'plane sections remain plane'. In other words, any deformation due to shear across the section is not accounted for (no shear deformation). During the relaxation experiment the curvature of the neutral axis of beam $\kappa$ remains constant over time, such that the bending moment $M_B$ continuously decreases. This case describes the relaxation of bending stress, assuming that the flexure deformation of beam does not alter in time.

Let $\sigma(z,t)$ is uniaxial stress in the beam in the direction of beam axis. The total strain in any instant of the time is $\varepsilon(z,t)$; is the sum of the elastic and the creep components of the strain:

(32)   $\varepsilon = \varepsilon_e + \varepsilon_c$.

The elastic component of shear strain is

(33)   $\varepsilon_e = \sigma / E$,

where $E$ is the shear modulus and $\varepsilon_c(z,t)$ is the creep component of normal strain.
Consider creep under constant in time total strain

(34)   $\varepsilon(z,t) = \varepsilon_e(z,t) + \varepsilon_c(z,t) \equiv \varepsilon_0(r)$.

The normal strain $\varepsilon_0(r) = \varepsilon(z, t=0)$ is the function of radius, but remains constant over time. The time differentiation of (34) leads to

(35)   $\dot{\varepsilon} = \dot{\varepsilon}_e + \dot{\varepsilon}_c \equiv 0$.

We exercise once again the common constitutive models of creep in bending state.

### 4.2. Bending moment relaxation for the material, that obeys Norton-Bailey Law

The Norton-Bailey law for a uniaxial state of stress reads

(36)   $\dot{\varepsilon}_c(z,t) = c_\sigma t^{k-1} \sigma^{m+1}$,

The substitution of material laws results in the ordinary differential equation for uniaxial stress

(37)   $\dfrac{\dot{\sigma}}{E} + c_\sigma t^{k-1} \sigma^{m+1} = 0$.

The initial condition for the equation (19) delivers the pure elastic shear stress in the initial moment

(38)   $\sigma(z, t=0) = \sigma_0(z)$.

For pure elastic bending the following initial distribution of stresses over the cross-section of the beam is valid:

$\sigma_0(z) = E \kappa z$,

where $\kappa$ is the bending curvature, which presumed to be constant over time and $z$ is the perpendicular distance to the neutral axis.
The solution of the ordinary differential equation (37) with initial condition (38) is

$$(39) \quad \sigma(z,t) = \left[\sigma_0^{-m}(z) + \frac{c_\sigma E m t^k}{k}\right]^{-1/m} = E\kappa \cdot \left[\frac{1}{z^m} + \frac{c_\sigma \kappa^m E^{m+1} m}{k} t^k\right]^{-1/m}.$$

The bending moment for the rectangular cross-section of width $B$ and height $H$ is the function of time.

$$M_B(t) = B \int_{-H/2}^{H/2} z\sigma(z,t)dz.$$

With the expression (39) we can calculate

$$(40) \quad M_B(t) = 2BE\kappa \int_0^{H/2} z \cdot \left[\frac{1}{z^m} + \frac{c_\sigma \kappa^m E^{m+1} m}{k} t^k\right]^{-1/m} dz.$$

Using the results of Appendix B ($J_p(a,m;X)$, case $p=1$), the integral in (40) could be expressed in terms of hypergeometric function

$$(41) \quad M_B(t) = 2BE\kappa \cdot J_1\left(\frac{c_\sigma \kappa^m E^{m+1} m t^k}{k}, m; \frac{H}{2}\right) =$$
$$= {}_2F_1\left(\frac{3}{m}, \frac{1}{m}; \frac{3+m}{m}; -\frac{c_\sigma \kappa^m E^{m+1} m t^k}{k}\left(\frac{H}{2}\right)^m\right) M_B^0,$$

where

$$M_B^0 = EBH^3\kappa/12$$

is the elastic bending moment at time $t=0$.

### 4.3. Bending moment relaxation for the material, that obeys Garofalo Law

We consider the now Garofalo law for uniaxial state of stress

$$(42) \quad \dot{\varepsilon}_c = c_\sigma t^{k-1} \sinh\left(\frac{\sigma}{\overline{\sigma}}\right).$$

The solution of the ordinary differential equation (37) with initial condition (38) for the Garofalo creep law leads to the expression of normal stress as function of coordinate $z$ and time $t$:

$$(43) \quad \sigma(z,t) = \overline{\sigma} \ln\left\{\tanh\left[\frac{Ec_\sigma}{2k\overline{\sigma}} t^k + \text{arctanh}\left(\exp\left(\frac{E\kappa}{\overline{\sigma}} z\right)\right)\right]\right\}.$$

For evaluation the formula for $I_1(a,b;X)$ from Appendix A is to be applied. With this formula the integral in (43) could be expressed in terms of polylogarithmic function

$$(44) \quad M_B(t) = 2B\overline{\sigma} I_1\left(\frac{Ec_\sigma}{2k\overline{\sigma}} t^k, \frac{E\kappa}{\overline{\sigma}}; \frac{H}{2}\right).$$

### 4.4. Bending moment relaxation for the material, that obeys Naumenko-Altenbach-Gorash Law

In this section we investigate the problem of the pure bending of a rectangular cross-section beam with a modified power law (stress range-dependent constitutive model) subjected to a bending moment. We assume the Naumenko-Altenbach-Gorash law for the state of uniaxial stress.

The substitution of modified power material law (9) results in the ordinary differential equation for uni-axial stress

$$(45) \quad \frac{\dot{\sigma}}{E} + \bar{\varepsilon} \cdot \left[ \frac{\sigma}{\bar{\sigma}} + \left( \frac{\sigma}{\bar{\sigma}} \right)^{m+1} \right] = 0.$$

When loaded by a bending moment, the beam bends so that the inner surface is in compression and the outer surface is in tension. The neutral plane is the surface within the beam between these zones, where the material of the beam is not under stress, either compression or tension. The solution of the ordinary differential equation (45) with initial condition (38) delivers the stress over the cross-section of the beam as the function of time and distance $z$ to neutral plane

$$(46) \quad \sigma(z,t) = \frac{z}{\left[ z^m \frac{m\bar{\varepsilon}E\varsigma - \bar{\sigma}}{\bar{\sigma}^{m+1}} + \frac{\varsigma}{E^m \kappa^m} \right]^{1/m}},$$

where

$$\varsigma = \exp\left( \frac{m\bar{\varepsilon}E}{\bar{\sigma}} t \right).$$

For calculation of the bending moment the formula for $J_p(a,m;X)$ from Appendix B is applied for the case $p=1$. With this formula the bending moment in the cross-section could be expressed in terms of hypergeometric function:

$$(47) \quad M_B(t) = {}_2F_1\left( \frac{3}{m}, \frac{1}{m}; \frac{3+m}{m}; \frac{(\kappa E H)^m}{2^m \bar{\sigma}^{m+1}} (\bar{\sigma} - m\bar{\varepsilon}E\varsigma) \right) \varsigma^{-1/m} M_B^0.$$

The curvature $\kappa$ of the beam remains constant in time. In the expressions (41), (44) and (47) the bending moment $M_B(t)$ is the function of time and continuously relaxes with time.

### 4.5. Basic constitutive equations for creep in bending

On the contrary, during the bending creep deformation the moment $M_B^0$ remains constant over time. The curvature $\kappa = \kappa(t)$ also continually increases. The elongation rate of the strip, which locates on the perpendicular distance $z$ to the neutral axis, is

$$\dot{\varepsilon}(t,z) = \dot{\kappa}z \qquad \text{for } -H/2 < z < H/2.$$

According to Norton-Bailey law (3) the shear stress due to creep is

$$\sigma_c = \left( \frac{\dot{\varepsilon}}{c_\sigma t^{k-1}} \right)^{\frac{1}{m+1}} = \left( \frac{\dot{\kappa}z}{c_\sigma t^{k-1}} \right)^{\frac{1}{m+1}}.$$

Performing the integration over the area of wire we get the bending moment due to creep

$$M_B^0 = B \int_{-H/2}^{H/2} z\sigma_c(z,t)dz = \frac{BH^2}{2} \frac{m+1}{3+2m} \left( \frac{H\dot{\kappa}}{2t^{k-1}c_\sigma} \right)^{\frac{1}{1+m}}.$$

Assuming that bending moment $M_B^0$ remains constant over time, from this equation the curvature time rate resolves as

$$(48) \quad \dot{\kappa} = \frac{2t^{k-1}c_\sigma}{H}\left(\frac{2M_B^0}{BH^2}\frac{3+2m}{m+1}\right)^{m+1}.$$

The rate of flexure $\dot{\kappa}$ remains constant, such that the bending radius linearly increases over time.

The results of this Section are applicable for springs, that overwhelmingly stressed by bending loads, like the leaf springs and torsion springs. Torsion springs may be of helix or spiral type. Accordingly the results of this Section could be instantly applied to estimate the effects of creep and relaxation of such springs.

### 4.6. Creep of disk springs

One important example of the springs with the uniaxial normal stress is the disk springs. In these springs dominates the state of bending of conical part.

For the calculation of disk springs the theory of the punched flat disks is usually applied. In this theory, the springs are considered to be flat circular disks, resting on their external rim and uniformly loaded along its internal rim [14]. The shape of the disks is not taken into account; nevertheless it has an influence on the shape of the suiting spring diagrams and on stress distribution. The load-deflection curve obtained by calculation is a straight line in this case. The maximum disks flattening load is identical to that obtained by the other calculation methods described hereinafter. The values of the stresses obtained by calculation are however lower than those obtained by calculation according the theory of the "conical cups". In the theory of the conical cups the spring disks is considered to be a conical cup supported on its external rim and uniformly loaded along its internal rim [15]. The calculation according Almen and Laszlo [16] starts from considering the spring disks to be a conical shaped cup, but neglecting the low radial stresses.

The method used in this article follows in general the method used by Almen and Laszlo for the description of static deformation of disk spring. The deformation hypothesis presupposes that the radial stresses are negligible and the cross section of spring does not distort, but rather that is merely rotates about the neutral point O.

Consider a sector of the disk, which is loaded by a force $P_z$, which assumed to be constant in time. Consider a sector and in it a strip $dx$ at location $x$ taking O as the origin (Fig. 17 of the cited article of Almen and Laszlo). Under the action of force $P_z$ the disk creeps and is deflected through a time dependent angle $\varphi$. For small axial deflection of disk $s$ the angle reads

$$\varphi = \frac{s}{A-B},$$

where $A$ and $B$ are respectively the outer and inner radii of disk.

This strip moves slowly into its new deformed position. The resultant tangential strain may be analyzed as the resultant of a radial displacement $\Delta r$ and a rotation $\varphi$. The first of these causes a uniform strain throughout the thickness of the disk $t_d$ if one neglects the small variation in distance to the center of the disk at various points of the section. The second results in a tangential bending strain which is zero in the neutral surface and maximum at the upper and lower surfaces. The tangential stresses produced by these two components of the strain cause a radial moment about point O which resists the moment created by the external and that of the deflected one forces.

The thickness of spring material $T$ is considered for brevity to be thin enough. This assumption allows neglecting the tangential stress due to bending. In other words, the influence of tangential bending strain is abandoned.

Calculating the tangential stress due to the radial displacement solely, we can use the expression for tangential strain in terms of rotation $\varphi$:

$$(49) \quad \varepsilon_\theta = \frac{x}{x-C}\left(\beta - \frac{\varphi}{2}\right)\varphi.$$

Here

$\beta = \dfrac{h}{A-B}$ is the initial cone angle of disk,

$C$ is the distance of neutral point O to the disk symmetry axis,

$h$ is the free height of disk, measured as the elevation of the truncated cone formed by either the upper or lower surface.

Under the creep conditions the rotation $\varphi = \varphi(t)$ and, consequently, tangential strain $\varepsilon_\theta = \varepsilon_\theta(t)$ are the functions of time. The tangential strain rate results form (49):

$$(50) \quad \dot{\varepsilon}_\theta \equiv \frac{d\varepsilon_\theta}{dt} = \frac{x}{x-C}\left[\left(\beta - \frac{\varphi}{2}\right)\dot{\varphi} + \left(-\frac{\dot{\varphi}}{2}\right)\varphi\right] \equiv \frac{x}{x-C}(\beta - \varphi)\dot{\varphi}.$$

Because it is assumed that the radial stresses are negligible, the tangential stress relates to tangential strain by means the uniaxial Norton- Bailey law

$$(51) \quad \dot{\varepsilon}_\theta(z,t) = c_\sigma t^{k-1} \sigma_\theta^{m+1}.$$

The tangential stress is the function of rotation angle $\varphi$ and its time derivative $\dot{\varphi}$

$$(52) \quad \sigma_\theta = \left[\frac{(\beta-\varphi)\dot{\varphi}}{c_\sigma t^{k-1}} \frac{x}{x-C}\right]^{\frac{1}{m+1}}.$$

The radial moment of the tangential forces in the section about point O is

$$(53) \quad dM_\theta = \sigma_\theta T x \sin(\beta - \varphi) dx\, d\theta.$$

Substituting in Equation (53) the Equation (52) and assuming the deflection as small,

$$\sin(\beta - \varphi) \cong \beta - \varphi,$$

we obtain the radial moment of the tangential forces in the section:

$$dM_\theta = (\beta - \varphi)\left[\frac{(\beta-\varphi)\dot{\varphi}}{c_\sigma t^{k-1}} \frac{x}{x-C}\right]^{\frac{1}{m+1}} T x\, dx\, d\theta.$$

Integrating from $x = C - A$ to $x = C - B$, we get the internal moment of the sector about point O:

$$(54) \quad M_\theta = (\beta - \varphi) 2\pi T \left[ \frac{(\beta - \varphi)\dot{\varphi}}{c_\sigma t^{k-1}} \right]^{\frac{1}{m+1}} \int_{C-A}^{C-B} \left[ \frac{x}{x-C} \right]^{\frac{1}{m+1}} x \, dx =$$

$$= (\beta - \varphi) 2\pi T \left[ \frac{(\beta - \varphi)\dot{\varphi}}{c_\sigma t^{k-1}} \right]^{\frac{1}{m+1}} L_{m+1}(A, B, C)$$

In the Equation (54) we make use of formulae from the Appendix C. The integral could be expressed analytically in terms of incomplete beta function.

The axial force of the disk spring $P_z$ is equal to the radial moment of the tangential forces divided by force arm $A - B$, such that

$$(55) \quad P_z = \frac{M_\theta}{A - B} = 2\pi T \left[ \frac{(\beta - \varphi)\dot{\varphi}}{c_\sigma t^{k-1}} \right]^{\frac{1}{m+1}} \frac{\beta - \varphi}{A - B} L_{m+1}(A, B, C) =$$

$$= 2\pi T \frac{L_{m+1}(A, B, C)}{A - B} \left[ \frac{(\beta - \varphi)^{m+2} \dot{\varphi}}{c_\sigma t^{k-1}} \right]^{\frac{1}{m+1}}.$$

The value of the $C$ in the last equation yet remains to be determined.

The sum of all forces action normal to the cross-section $P_\theta$ must be equal to zero. Only stresses due to radial displacement need to be considered. To calculate the sum of all forces action normal to the cross-section, we make use of the expression for tangential stress as the function of rotation angle $\varphi$ and its time derivative $\dot{\varphi}$:

$$(56) \quad P_\theta = \int_{C-A}^{C-B} \sigma_\theta T \, dx = T \left[ \frac{(\beta - \varphi)\dot{\varphi}}{c_\sigma t^{k-1}} \right]^{\frac{1}{m+1}} \int_{C-A}^{C-B} \left[ \frac{x}{x-C} \right]^{\frac{1}{m+1}} dx = 0.$$

Hence, the position of neutral point $C$ is to be determined from the equation

$$(57) \quad \int_{C-A}^{C-B} \left[ \frac{x}{x-C} \right]^{\frac{1}{m+1}} dx \equiv K_{m+1}(A, B, C) = 0.$$

The integral in the Equation (57) allows again the representation in terms of incomplete beta functions using the formulae from the Appendix C. Further simplification and closed form solution of the Equation (57) for an arbitrary $m$ seems to be impossible. For $m = 0$ one get the known solution

$$(58) \quad C = \frac{B - A}{\ln(B/A)}.$$

The appropriate approximate solution for sufficiently large $m$ delivers the expression

$$(59) \quad C = \frac{A + B}{2}.$$

For constant $A, B, C$ the expression

$$K_{m+1}(A, B, C)$$

is the function of creep exponent $m$.

On one side, for higher values of exponent the first function $K_{m+1}(A,B,(A+B)/2)$ asymptotically tends to zero axis.

On the other side, for higher values of exponent the second function $K_{m+1}(A,B,(B-A)/\ln(B/A))$ gradually deviates from zero.

For illustration two functions

$$K_{m+1}(A,B,(B-A)/\ln(B/A))$$

and

$$K_{m+1}(A,B,(A+B)/2)$$

are drawn on the Fig. 3. For lower values of $m$ the solution (58) is delivers better approximation for the equation

$$K_{m+1}(A,B,C)=0.$$

Two functions

$$L_{m+1}(A,B,(B-A)/\ln(B/A))$$

and

$$L_{m+1}(A,B,(A+B)/2)$$

are drawn on the Fig. 4. The approximate solution (58) leads to somewhat higher values of function $L_{m+1}$, than the solution (59).

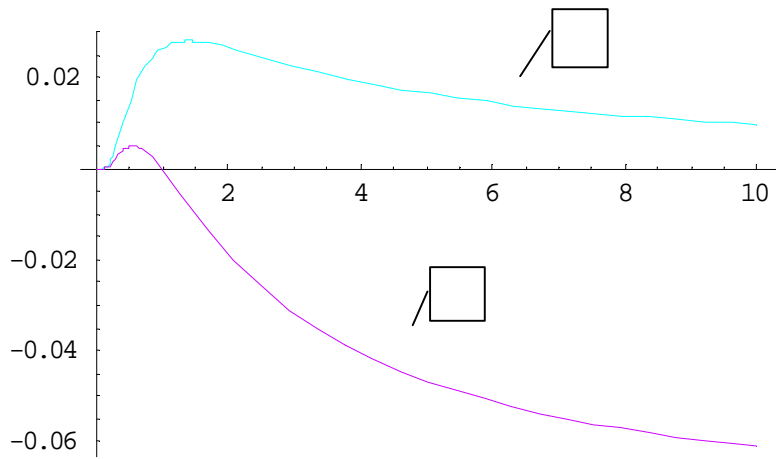

**Fig. 3 Two functions** $K_{m+1}(A,B,(B-A)/\ln(B/A))$ **and** $K_{m+1}(A,B,(A+B)/2)$ **for different exponents** $m$

With the approximate solution (59) for the distance of neutral axis to center we get the final expression of the spring force as the function of rotation angle $\varphi(t)$:

$$(60) \quad P_z = \left[\frac{(\beta-\varphi)^{m+2}\dot{\varphi}}{c_\sigma t^{k-1}}\right]^{\frac{1}{m+1}} \frac{2\pi T}{A-B} L_{m+1}\left(A,B,\frac{A+B}{2}\right).$$

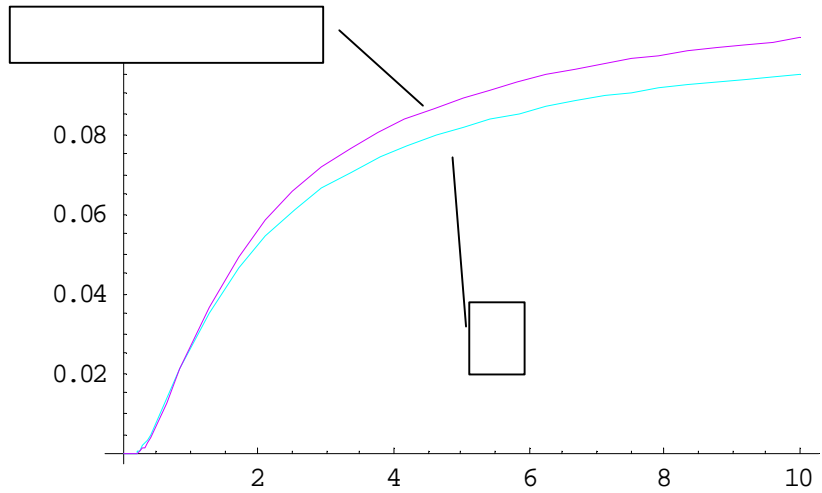

**Fig. 4 .Two functions** $L_{m+1}(A,B,(B-A)/\ln(B/A))$ **and** $L_{m+1}(A,B,(A+B)/2)$ **as functions of exponent** $m$

If the spring force remains constant over time, the rotation angle $\varphi(t)$ is the solution of an ordinary differential equation

(61) $\quad t^{1-k}(\beta-\varphi)^{m+2}\dot{\varphi} = \tilde{P}$ , $\varphi(0) = \varphi_0$ ,

Where

$$\tilde{P} = c_\sigma \left[ \frac{(A-B)P_z}{2\pi T L_{m+1}\left(A,B,\frac{A+B}{2}\right)} \right]^{m+1}$$

$\varphi_0$ the initial rotation angle of the cone due to elastic deformation of the spring at time moment $t = 0$.

The solution of the equation (61) delivers the rotation angle as function of time:

(62) $\quad \varphi(t) = \beta - \left[ (\beta-\varphi_0)^{m+3} - \frac{(m+3)\tilde{P}t^k}{k} \right]^{\frac{1}{m+3}}$.

At the critical moment

$$t_{crit} = \left( \frac{k(\beta-\varphi_0)^{m+3}}{(m+3)\tilde{P}} \right)^{1/k}$$

the cone angle of disk vanishes and spring turns to be a flat disk, such that $\varphi(t_{crit}) = \beta$.
The axial deflection of disk is the function of time:

$$s(t) = \varphi(t)(a-b).$$

For example, we calculate the creep of the disk spring with the following parameters:

$$A = 25\,mm, B = 12.25\,mm, T = 3\,mm, h = 1.1\,mm.$$

The Figures 5 and 6 demonstrate the influence of parameter $\bar{\sigma}$ on the creep behavior of disk spring. The initial rotation of spring cone is $\varphi_0 = 3\beta/4$. In other words, this means that the initial spring travel is $s_0 = 3h/4 = 0.825\,mm$. The following material parameters were used for calculation:

$$E = 200\,GPa,\quad m = 4,\ k = 1, \nu = 0.254, \bar{\varepsilon} = 10^{-24}\,\sec^{-1}.$$

On the Fig.5 the creep curves for three different material parameters are shown:

$$\bar{\sigma} = 1600\,MPa,\ \bar{\sigma} = 1800\,MPa\ \text{and}\ \bar{\sigma} = 2000\,MPa.$$

The creep curves represent the cone angles as functions of time. The relaxation happens slower for the materials with higher values of parameter $\bar{\sigma}$.

On the Fig.6 are shown the disk heights under constant load as function of time for three different material parameters. The cone angles of disk vanish for three different critical moments. Sooner or later - depending on material parameter - the springs turn to be a flat disk. The disk flattening happens evidently at higher time moments for the materials with elevated parameters $\bar{\sigma}$.

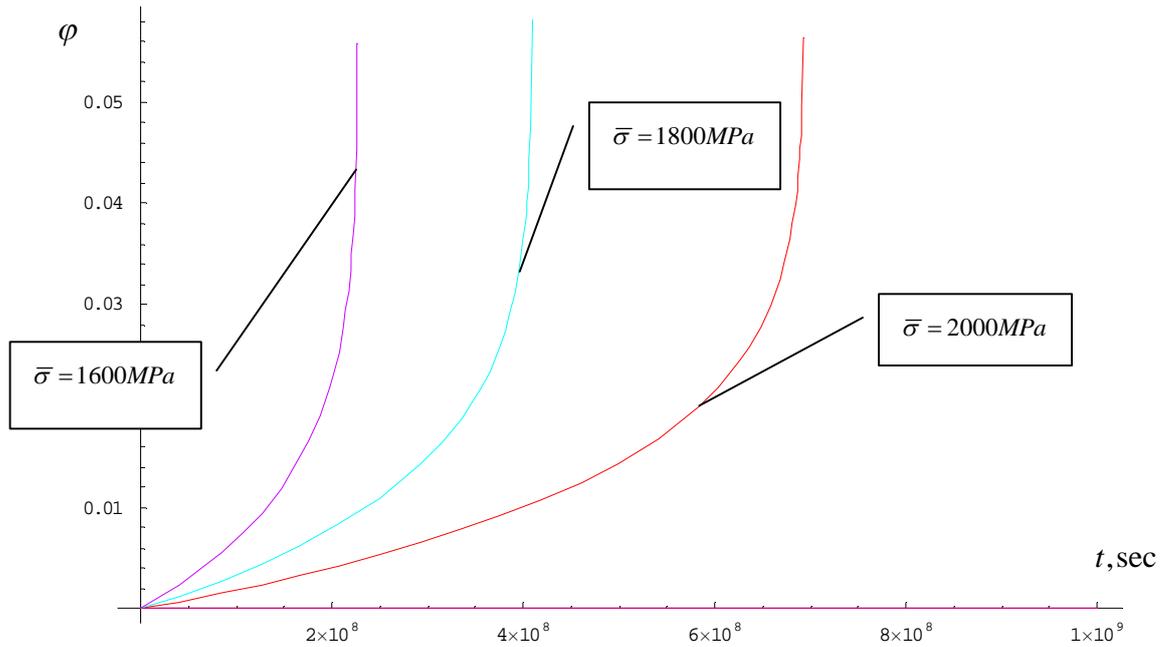

**Fig. 5. Cone angles of disk springs as functions of creep time for three different material parameters**

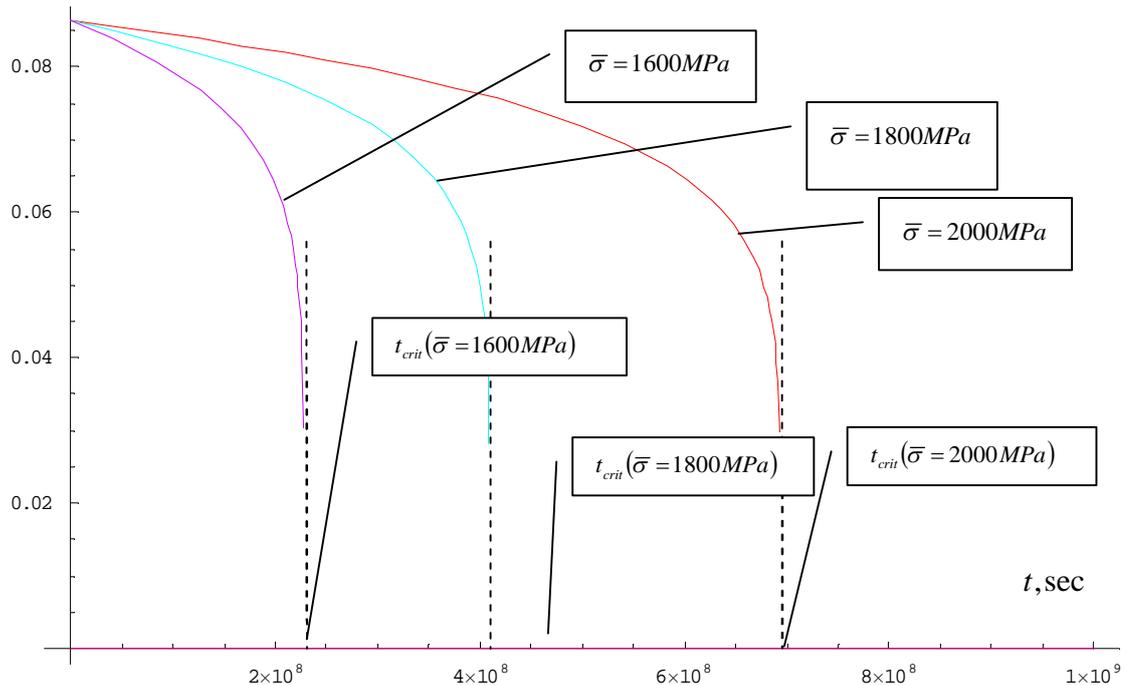

**Fig. 6. Disk heights as functions of creep time for three different material parameters**

## 5. Conclusion

The aim of this paper is to examine further the detailed behavior of simple structures with common creep laws. The relaxation of stresses for Norton-Bailey, Garofalo and Naumenko-Altenbach-Gorash constitutive models was studied for basic elements in torsion and bending. The structures examined are elementary - a beam in bending, a rod in torsion and helical and disk springs - but have long been used to demonstrate the basic characteristics of nonlinear creep. As the bending and torsion frequently occur in structural members, the results are immediately applicable. The application of the solutions allows accurate analytic description of creep and relaxation of practically important problems in mechanical engineering. Following the procedure we establish closed form solutions for creep and relaxation in helical, leaf and disk springs.

The closed form solutions with commonly accepted creep models allow a deeper understanding of such a constitutive model's effect on stress and deformation and the implications for high temperature design.

**Appendix A. Integrals with polylogarithm**

The weighted integrals of the function

$$f(x) = \ln(\tanh(a + \operatorname{arctanh}(\exp(bx))))$$

are:

$$I_0(a,b;X) \equiv \int_0^X f(x)dx = X\ln(\tanh(a)) + \frac{1}{b}(\Lambda_2 - M_2 + \mu_2 - \lambda_2),$$

$$I_1(a,b;X) \equiv \int_0^X f(x)xdx =$$
$$= \frac{1}{6b^2}\left[\pi^2 \ln(\coth(a)) + \ln^3(\coth(a)) + 3b^2 X^2 \ln(\tanh(a))\right] +$$
$$+ \frac{1}{b^2}(M_3 - \Lambda_3) - \frac{X}{b}(M_2 - \Lambda_2),$$

$$I_2(a,b;X) \equiv \int_0^X f(x)x^2 dx = \frac{X^3}{3}\ln(\tanh(a)) +$$
$$+ \frac{2}{b^3}(\Lambda_4 - M_4 + \mu_4 - \lambda_4) + \frac{2X}{b^2}(M_3 - \Lambda_3) - \frac{X^2}{b}(M_2 - \Lambda_2).$$

The following abbreviations are used:

$$M_k = \operatorname{Li}_k(-\coth(a)\exp(bX)),$$
$$\Lambda_k = \operatorname{Li}_k(-\tanh(a)\exp(bX)),$$
$$\mu_k = M_k\big|_{X=0} \equiv \operatorname{Li}_k(-\coth(a)),$$
$$\lambda_k = \Lambda_k\big|_{X=0} \equiv \operatorname{Li}_k(-\tanh(a)).$$

In these expressions is $\operatorname{Li}_k(z)$ the polylogarithm (also known as Jonquière's function) of order $k$ and argument $z$. For the derivation of formulas of this Appendix we use the expression of polylogarithm as the integral of the Bose–Einstein distribution:

$$\operatorname{Li}_k(z) = \frac{1}{\Gamma(k)}\int_0^\infty \frac{x^{k-1}}{z^{-1}\exp(x) - 1}dx.$$

This integral converges for Re($k$) > 0 and all $z$ except for $z$ real and $\geq 1$.

**Appendix B. Integrals with hypergeometric function**

The weighted integrals of the function

$$g = \left(a + x^{-m}\right)^{-1/m},$$

are

$$J_p(a,m;X) \equiv \int_0^X x^p g(x)\,dx =$$
$$= {}_2F_1\left(\frac{1}{m}, \frac{2+p}{m}; \frac{2+p+m}{m}; -aX^m\right)\frac{X^{2+p}}{2+p}, \qquad p \geq 0.$$

For some cases the integrals could be expressed in terms of elementary functions:

$$J_1(a,1;X) = -\frac{X}{a^2} + \frac{X^2}{2a} + \frac{\ln(1+aX)}{a^3},$$

$$J_2(a,1;X) = \frac{X}{a^3} - \frac{X^2}{2a^2} + \frac{X^3}{3a} - \frac{\ln(1+aX)}{a^4},$$

$$J_1(a,2;X) = \frac{X\sqrt{1+aX^2}}{2a} - \frac{\arcsin(X\sqrt{A})}{2a^{3/2}},$$

$$J_2(a,2;X) = \frac{2}{3a^2} - \frac{2\sqrt{1+aX^2}}{3a^2} + \frac{X^2\sqrt{1+aX^2}}{3a},$$

$$J_1(a,3;X) = -\frac{1}{2a} + \frac{\left(1+aX^3\right)^{2/3}}{2a},$$

$$J_2(a,4;X) = -\frac{1}{3a} + \frac{\left(1+aX^4\right)^{3/4}}{3a}.$$

**Appendix C. Integrals with incomplete beta function**
The weighted integrals of the function

$$\left(\frac{|c-r|}{r}\right)^{1/n}$$

are:

$$K_n(a,b,c) \equiv \frac{1}{c}\int_a^b \left(\frac{|c-r|}{r}\right)^{1/n} dx = \frac{1}{c}\left[\int_a^c \left(\frac{c-r}{r}\right)^{1/n} dx + \int_c^b \left(\frac{r-c}{r}\right)^{1/n} dx\right] =$$

$$= \frac{1}{c}\lim_{\varepsilon \to +0}\left[\int_a^c \left(\frac{c-r+\varepsilon}{r}\right)^{1/n} dx + \int_c^b \left(\frac{r-c+\varepsilon}{r}\right)^{1/n} dx\right] =$$

$$= \frac{(-1)^{-1/n}\pi}{n}\left(i + \cot\left(\frac{\pi}{2n}\right)\right) - B\left(\frac{a}{c}; \frac{n-1}{n}, \frac{n+1}{n}\right) - (-1)^{1/n} B\left(\frac{b}{c}; \frac{n-1}{n}, \frac{n+1}{n}\right)$$

and

$$L_n(a,b,c) \equiv \frac{1}{c^2}\left[\int_a^b \left(\frac{|c-r|}{r}\right)^{1/n}(c-r)dx\right] =$$

$$= \frac{1}{c^2}\left[\int_a^c \left(\frac{c-r}{r}\right)^{1/n}(c-r)dx + \int_c^b \left(\frac{r-c}{r}\right)^{1/n}(c-r)dx\right] =$$

$$= \frac{1}{c^2}\lim_{\varepsilon \to +0}\left[\int_a^c \left(\frac{c-r+\varepsilon}{r}\right)^{1/n}(c-r)dx + \int_c^b \left(\frac{r-c+\varepsilon}{r}\right)^{1/n}(c-r)dx\right] =$$

$$= \frac{i\pi}{\exp\left(\frac{i\pi}{n}\right)-1}\frac{1+n}{n^2} + B\left(\frac{a}{c}; \frac{2n-1}{n}, \frac{n+1}{n}\right) - B\left(\frac{a}{c}; \frac{n-1}{n}, \frac{n+1}{n}\right) +$$

$$+ (-1)^{-1/n}\left[B\left(\frac{b}{c}; \frac{2n-1}{n}, \frac{n+1}{n}\right) - B\left(\frac{b}{c}; \frac{n-1}{n}, \frac{n+1}{n}\right)\right].$$

In these expressions

$$B(x; p, q) = \int_0^x z^{p-1}(1-z)^{q-1} dz$$

is the incomplete beta function [**17**]**.**
It is given in terms of hypergeometric functions by

$$B(x; p, q) = {}_2F_1(p, 1-q; p+1; x)\frac{x^p}{p}.$$

# Addendum to "Relaxation and Creep in Twist and Flexure"

Kobelev V.[1]

## 1. Generalized Norton-Bailey Law

When a plastic material is subjected to a constant load, it deforms with a variable strain rate. The initial strain is roughly predicted by its stress-strain modulus. The material will continue to deform gradually with time, until yielding or rupture causes failure.

The models adequately describe the primary and secondary creep stage from constant load/stress uniaxial tests where creep rate is nearly constant.

The primary region is the early stage of loading when the creep rate decreases rapidly with time. The creep deformation of soft metals at constant temperature and stress grows in time according to a power law with an exponent. According to the Andrade law (Andrade 1910, 1914), the creep strain rate in the primary region could be described by the relation

$$(1) \quad F_I(\sigma_{eff}, t) = \varepsilon_I \left(\frac{t}{T}\right)^{k-1} \left(\frac{\sigma_{eff}}{\bar{\sigma}}\right)^{m+1} \quad \text{or}$$

$$F_I(\sigma_{eff}, t) = \varepsilon_I \exp\left[-\left(\frac{t}{T}\right)^k\right] \left(\frac{\sigma_{eff}}{\bar{\sigma}}\right)^{m+1}.$$

Then it reaches a steady state which is called the secondary creep stage. The creep strain rate in the secondary region defines the Norton-Bailey law (Betten, 2008)

$$(2) \quad F_{II}(\sigma_{eff}, t) = \varepsilon_{II} \left(\frac{\sigma_{eff}}{\bar{\sigma}}\right)^{m+1}.$$

Here

$$\varepsilon_I, \varepsilon_{II}, \bar{\sigma}, t, T, m, k$$

are the experimental constants for the first and secondary stages of creep. The other notations follow the cited Article (Kobelev, 2014).

The generalized Norton-Bailey law represents the experimentally acquired creep laws. The isotropic stress function reads for both primary and secondary case

$$(3) \quad F(\sigma_{eff}, t) = F_I(\sigma_{eff}, t) + F_{II}(\sigma_{eff}, t).$$

Function (3) could be represented as the product of two functions, which depend correspondingly on time and on stress:

$$(4) \quad F(\sigma_{eff}, t) = h(t) s(\sigma_{eff}).$$

The creep law (4) pronounces the continuous transition between the primary and secondary creep regimes. Primary creep occurs at the beginning of the creep test and is characterized first by a high strain rate. The creep rate decelerates gradually to a constant value denoting the beginning of secondary creep. This behavior corresponds to the experimentally acquired creep data (Nabarro, de Villers, 1995) and (Es-Souni, 2000). However, the continuous transition between creep regimes leads to some mathematical difficulties. The regions of primary and secondary creep are frequently being separated for reasons of mathematical simplification (Nezhad, O'Dowd (2012, 2015).
The derivation of closed form solution (Kobelev, 2014) is applied in this Manuscript to the generalized creep law (4). The most general form of creep law with continuous transition between the primary and secondary creep regimes permits the closed form solutions because of simple geometry of considered stress states.
Particularly, for the generalized Norton-Bailey law the functions in (4) are

$$h(t) = \varepsilon_I \left(\frac{t}{T}\right)^{k-1} + \varepsilon_{II}, \quad \text{or}$$

$$h(\sigma_{eff}, t) = \varepsilon_I \exp\left[-\left(\frac{t}{T}\right)^k\right] + \varepsilon_{II},$$

$$s(\sigma_{eff}) = \left(\frac{\sigma_{eff}}{\overline{\sigma}}\right)^{m+1}.$$

---

[1] kobelev@imr.mb.uni-siegen.de, University of Siegen, D-57076, Siegen, Germany

The experimentally acquired creep laws demonstrate normally the behavior, proclaimed by the Eq. (4).
The mathematical treatment is not restricted by use ordinary derivatives. Well known, that the use of fractional calculus generalized the conventional mechanical models. The Newton element of common creep laws is replaced by the Scott–Blair element (of order $\alpha$). The immediate way to introduce fractional derivatives in the constitutive equation for creep (Mainardi, Spada, 2011) is the substitution the first derivative with a fractional Caputo derivative

$$D^\alpha f(t) = \int_0^t \frac{(t-\xi)^{-\alpha}}{\Gamma(1-\alpha)} \frac{df(\xi)}{d\xi} d\xi$$

of order

$$0 < \alpha < 1.$$

The limit case of $\alpha = 1$ provides the ordinary derivative.
The additional constant $\alpha$ is to be chosen to better fit of the measured creep strain as the function of time. Consequently, the use of fractional derivatives facilitates the mathematical approximations of the experimentally acquired creep laws.

## 2. Relaxation of twisted rods

### 2.1. Basic constitutive equations for relaxation

Consider the relaxation problem for a rod with circular cross-section with the radius *R* under the constant twist. The shear stress $\tau$ over the cross-section of the twisted rod is the function of time *t* and radius $0 < r \leq R$ (Kobelev, 2014, Eq. 19). With

$$\sigma_{eff} = \sqrt{3}\tau,$$

$$\bar{\tau} = \bar{\sigma}/\sqrt{3},$$

$$\bar{\gamma} = \sqrt{3}\bar{\varepsilon}$$

and shear modulus $G$ the shear stress reads as

$$(5) \quad \tau(r,t) = \left[\tau_0^{-m}(r) + \frac{\sqrt{3}Gm}{\bar{\tau}^{m+1}} h_\alpha(t)\right]^{-1/m},$$

where

$$(6) \quad h_\alpha(t) = \begin{cases} D^{-\alpha} h(t), & 0 < \alpha < 1, \\ \int_0^t h(p)dp, & \alpha = 1. \end{cases}$$

is the fractional integral of the order $\alpha$ for $0 < \alpha < 1$ and an ordinary integral in the case $\alpha = 1$.

The torque $M_T(t)$ is the function of time

$$(7) \quad M_T(t) = 2\pi \int_0^R r^2 \tau(r,t) dr.$$

The integral on the right side of (7) could be expressed in terms of hypergeometric function for an appropriate value of power $p$. The integrand in (7) depends on radius of power 2, such that $p = 2$. The formula (Kobelev, 2014, Appendix B) being applied in the case $p = 2$, delivers the relaxation function for torque

$$(8) \quad \frac{M_T(t)}{M_T^0} = {}_2F_1\left(\frac{1}{m}, \frac{2+p}{m}; \frac{2+p+m}{m}; -\sqrt{3}m\left(\frac{\tau_{max}}{\bar{\tau}}\right)^{m+1} \frac{G}{\tau_{max}} h_\alpha(t)\right).$$

The initial torque at the moment $t = 0$ is

$$M_T^0 = G\pi\theta R^4/2.$$

The angle of twist of the rod with the length $L$ is

$$\alpha = L\theta.$$

The spatial twist rate

$$\theta \equiv \alpha/L = \tau_{max}(GR)^{-1}$$

is constant in time and depends solely upon the shear stress on the outer surface of the rod $\tau_{max}$. Consequently, the relaxation function is solely the function of the shear stress on the surface $\tau_{max}$ and material parameters.

### 3.    Relaxation of beams in bending

**3.1. Basic constitutive equations for relaxation in bending**

Consider the problem of stress relaxation in the pure bending of a rectangular cross-section (BxH) beam. In the applied Euler-Bernoulli theory of slender beams subjected to a bending moment $M_B$, such that any deformation due to shear across the section is not accounted for.

During the relaxation experiment the curvature of the neutral axis of beam remains constant over time, such that the bending moment $M_B$ continuously decreases. This case describes the relaxation of bending stress, assuming that the flexure deformation of beam does not alter in time.

The stress in the beam in the direction of beam axis $\sigma(z,t)$ is the function of time $t$ and coordinate

$$-H/2 \leq z \leq H/2 \ .$$

The solution of the fractional differential equation for bending (Kobelev, 2014, Eq. 39) with

$$\sigma_{eff} = \sigma$$

and Young modulus $E$ reads

(9) $\sigma(z,t) = \left[ \sigma_0^{-m}(z) + \frac{E m}{\bar{\sigma}^{m+1}} h_\alpha(t) \right]^{-1/m}.$

The bending moment for the rectangular cross-section of width $B$ and height $H$ is the function of time

(10) $\quad M_B(t) = B \int_{-H/2}^{H/2} z \sigma(z,t) dz.$

The integral in (10) could be expressed in terms of hypergeometric function (Kobelev, 2014, Appendix B).

The integrand in (10) depends on $z$ of power 1, such that the formula of the Appendix B should be applied for $p = 1$.

From this integral the relaxation function for internal bending moment reads:

(11) $\quad \dfrac{M_B(t)}{M_B^0} = {}_2F_1\left( \dfrac{1}{m}, \dfrac{2+p}{m}; \dfrac{2+p+m}{m}; -m\left(\dfrac{\sigma_{max}}{\bar{\sigma}}\right)^{m+1} \dfrac{E}{\sigma_{max}} h_\alpha(t) \right).$

where

$$M_B^0 = EBH^3 \kappa / 12$$

is the elastic bending moment at time

$$t = 0.$$

The curvature of rod

$$\kappa = 2\sigma_{max}(EH)^{-1}$$

depends solely on the normal stress at the uppermost edge of the beam $\sigma_{max}$. Once again, the relaxation function is the function of the maximal normal stress $\sigma_{max}$ and material parameters.

## 4. Solutions for other creep laws

The method could be applied for definite other stress functions $s(\sigma_{eff})$ in the creep law (4). The isotropic stress function for Garofalo creep law (Garofalo, 1963) reads

$$s(\sigma_{eff}) = \sinh(\sigma_{eff}/\bar{\sigma}),$$

$$F(\sigma_{eff}, t) = h(t)\sinh(\sigma_{eff}/\bar{\sigma}).$$

generalizes the Eqs. (24) and (44) in (Kobelev, 2014) with the arbitrary time function $h(t)$. In the same way the Naumenko-Altenbach-Gorash creep law (Naumenko et al., 2009)

$$s(\sigma_{eff}) = (\sigma_{eff}/\bar{\sigma}) + (\sigma_{eff}/\bar{\sigma})^{m+1},$$

$$F(\sigma_{eff}, t) = h(t)\left[(\sigma_{eff}/\bar{\sigma}) + (\sigma_{eff}/\bar{\sigma})^{m+1}\right]$$

delivers the closed form solutions for the relaxation functions the Eqs. (28) and (47) in (Kobelev, 2014).

## 5. Conclusion

This Manuscript introduces the new expression for creep law. The aim of this paper is to examine further the detailed behavior of simple structures with generalized creep law with separable variables. The new expression is based on the experimental data and unifies the primary, secondary and tertiary regions of creep curve. The relaxation functions for bending and torsion depend only on the maximal stress in the cross-section, which occurs on the outer surface.

Nezhad H.Y., O'Dowd N.P. (2015) Creep relaxation in the presence of residual stress - Engineering Fracture Mechanics, 138, 250-264